\documentstyle{elsart}

\begin{document}
\begin{frontmatter}
\title{Transmission properties of the X-ray window for the SIXA
 spectrometer}

\author{T. Tikkanen and J. Huovelin}
\address{Observatory and Astrophysics Laboratory,
P.O. Box 14 (T\"ahti\-tornin\-m\"aki),
 FIN-00014 University of Helsinki,
 Finland}

\begin{abstract}
The ultrathin X-ray entrance window for the spaceborne spectrometer SIXA
was characterised using synchrotron radiation.
X-ray absorption fine structure near the absorption edges of
the constituting elements (aluminium, carbon, oxygen and nitrogen)
 was measured.
Large scale positional variations were also studied.
In addition, the opacity of the window
was tested by observing the effect of
 photocurrent in the detector generated by light leaked through the
 window.
\end{abstract}
\end{frontmatter}

\section{Introduction}

SIXA (Silicon X-Ray Array) spectrometer \cite{Vilhu}
 is a focal plane instrument of the SODART X-ray telescope onboard the
 Spectrum-X-Gamma satellite.
SIXA is an array of 19 discrete Si(Li) detector elements which collect
 X-rays in the range 0.5--20~keV with an energy resolution of about
 200~eV at 6~keV\@. The elements are circular with an active diameter of
 9.2~mm and they are arranged in a hexagonal pattern with a centreline
 distance of 12~mm. The detector crystals are kept at a temperature of
 about 120~K by a passive cooling system. X-rays focussed by the
 telescope enter the detector through an entrance window which forms a
 part of the cooler shield.

The X-ray entrance window for SIXA was fabricated by Metorex
 International Oy. A specific advantage of this window construction
 \cite{Viitanen} is that a polyimide mesh is utilised to support an
 ultrathin polyimide membrane. Thus the active area of the window
 (70~mm in diameter) is shadowed by the support structure only at the
 lowest X-ray energies. In addition to being transparent in X-rays, the
 window is required to be tight against heat leakage,
opaque to light from near~IR to far~UV
and compatible with space environment.
The window comprises two separate units, both coated with aluminium.

Because the window absorbs a considerable fraction of the
 softer X-rays, determination of its X-ray transmittance is an
important part of the detector characterisation. Transmission properties
can be fairly well predicted from tabulated data for the constituent
 elements over the energy range of SIXA except for the regions
above the K~absorption edges of aluminium and oxygen. Transmission in
 these regions is affected by X-ray absorption fine structure
 (XAFS) which depends on the chemical structure of the materials.

We applied synchrotron radiation from the electron storage ring BESSY
to investigate the soft X-ray transmission properties. In addition to
the energy dependence of the transmittance, we studied the dependence
 on the position on the window surface. As SIXA consists of large
 elements which do not have any spatial resolution, only large scale
 positional variations were of interest.
Furthermore, we studied the opacity to light, which is required
to suppress the degradation of the X-ray detection performance
by photocurrent in the Si(Li) crystals.

\section{X-ray window units}

In the fabrication process a layer of BPDA-PPD PI-2610 polyimide was
 first made by spin casting a low viscosity polyamic acid resin onto a
silicon wafer and baking the resin for conversion to polyimide. Next the
 support grid was formed of photosensitive PI-2732 polyimide which was
patterned into a hexagonal honeycomb structure with a 500~$\mu$m pitch.
Finally, the silicon substrate was removed and a 30~nm layer of
aluminium was sputtered on both sides.
Thus one window unit consists of a thin layer of polyimide, a
 polyimide support grid, and a layer of aluminium on each surface.
The chemical composition of the polyimide is
H\raisebox{-0.5ex}{\small 10}C\raisebox{-0.5ex}{\small
 22}N\raisebox{-0.5ex}{\small 2}O\raisebox{-0.5ex}{\small 4}
and its bulk density is about 1.4~g/cm$^3$.
The density of the thin film polyimide is about 5\% higher.

A set of window units was manufactured and the best ones were selected
 for the flight model (FM) and the flight spare model (FSM) of SIXA\@.
The polyimide parameters of these as well as a few other units used for
 calibration are given in Table~\ref{tab:pi}.
Nominal aluminium thickness is 60~nm for all units in the table.
Native oxide layers of thickness of about 3~nm
are formed on the aluminium coatings.
Thicknesses of the polyimide membranes
were measured with a calibrated profilometer and they
varied from 237 to 275~nm. However, it was found later that
a soft coating on a hard substrate, such as polyimide on silicon,
is compressed during the measurement \cite{Harvela}.
The tabulated values have been corrected to account for the error.
The shadowing figures are averaged, which is sufficient because
the detector elements are large compared to the mesh
(the active area covers 300 hexagons).

\section{X-ray transmission}

\subsection{Experiment}

Window units 19 and 14 were used for the X-ray transmission
 measurements. The main purpose of the experiment was to measure the
 X-ray absorption fine structures which would be used to calculate the
 transmission properties of the FM and FSM window units. As all window
 units are composed of similar layers produced by the same process, they
 are likely to have similar XAFS\@.

The soft X-ray transmittance of the two window units was studied at the
 SX700 plane grating monochromator beamline \cite{Scholze} of the
radiometry laboratory of the Physikalisch-Technische Bundesanstalt (PTB)
 which is located at the BESSY synchrotron facility. Control over the
 position of the beam was provided by a movable sample holder.
The beam covered the area of about 5~hexagons of the honeycomb grid
 structure with its full width at tenth maximum (1.5~mm horizontal and
0.8~mm vertical).
Consequently, the fraction $S$ of the beam shaded by the grid
was affected by the precise positioning of the beam,
and the values of $S$ during separate measurements varied
about the average shadowing given in Table~\ref{tab:pi}.

Transmittance as a function of energy was obtained by scanning photon
energies from 60~eV to 1800~eV with steps varying from 0.1~eV to 10~eV,
depending on the expected detailed features in the transmittance curve.
The beam was positioned at the centre. The results are plotted in
 Fig.~\ref{fig:e-scan}.
The similarity of the two window units throughout the measured range,
 including the absorption edges, suggests that XAFS is indeed
 identical for all SIXA window units.
The near-edge structures of the polyimide constituents resemble those
 reported for other windows manufactured by Metorex \cite{Bavdaz},
while XAFS of aluminium is different which is natural because there are
no AlN layers in our windows.

Dependence on position was studied by performing line scans in the $x$
direction across the window surface at a few settings of the $y$
 coordinate. This was repeated at four discrete energies slightly below
 the K~edges of the elements C, N, O and Al. The scan step was 2~mm.
Transmission along the scanned lines is shown in Fig.~\ref{fig:x-scan}.
It was originally measured in arbitrary units and later scaled to the
transmittances of the energy scans by fitting a second order polynomial
to the data from the scan along $y=0$ and setting the value of the
 polynomial in the centre (its $x$ coordinate being that obtained in the
analysis in section \ref{x-scan}) equal to the energy scan transmittance
 at the respective energy.
The transmittance grows with the radial distance from the centre
of the window.

\subsection{Analysis of energy scans} \label{e-scan}

The model of the transmittance of a window unit is
\begin{equation} \label{model}
T = \exp \left[ - \sum_{j=1}^3
		    \left( \frac{\mu}{\rho} \right)_j \rho_j t_j
	 \right]
    \left\{ 1 - S + S \exp \left[ - \left( \frac{\mu}{\rho} \right)_4
				    \rho_4 t_4
			   \right]
    \right\} ,
\end{equation}
where the sum goes over the three thin film materials
 (aluminium, oxide and polyimide) and index 4 refers to the grid.
This equation, with the experimental data for $T$, was applied to
 calculate the mass attenuation coefficients ($\mu/\rho$) wherein the
 fine structures are incorporated. The transmittance of the FM window
 units was then calculated using the model and the
values of Table~\ref{tab:pi} for $S$ and the layer thicknesses $t_j$.

The experimental data is compared to the model (Eq.~\ref{model} with
 absorption data from ref.~\cite{Henke} and window
 parameters from Table~\ref{tab:pi}) in Fig.~\ref{fig:fit}a. The
 difference between the residuals of the two window units is plotted in
 Fig.~\ref{fig:fit}b. The residuals exhibit considerable structure above
 each absorption edge with very similar shape for the two units. Most of
 the residuals can therefore be attributed to XAFS which can extend
 several hundred eV above the K~edges. The residuals in the range
 600--1550~eV are due to the deviation of the actual beam shadowing $S$
 during the energy scans from the average values in Table~\ref{tab:pi}.
 Measurement errors and errors in the $t_j$ data of Table~\ref{tab:pi}
 are also partly responsible for the residuals. Because the contribution
 to the residuals from the errors in the $t_j$ is smaller than the
 features arising from XAFS, more accurate values for the $t_j$ than
 those in Table~\ref{tab:pi} can not be abstracted from the data.
 Moreover, derivation of layer thicknesses from transmission data yields
 poor results in general, because the dependence of $\mu/\rho$ on the
X-ray energy is very similar for all materials outside the XAFS regions.
 In consequence, when fitting a function of the form of Eq.~\ref{model}
 to the data, the thicknesses are strongly correlated and their best-fit
 values have very large error ranges.

However, the difference between the residual curves of the two units
can be attributed to relative errors in the $t_j$ between the units.
We reduced the relative errors with a set of relative corrections
 $\Delta t_j$, where the corrected values of the layer thicknesses
 $t_{j,19}$ and $t_{j,14}$ of the two units are
 $t_{j,19} + \Delta t_j$ and $t_{j,14} - \Delta t_j$.
First we minimised the $L^1$ norm of the difference in the range
60--530~eV with the beam shadowing and the corrections of the thin film
 thicknesses as free parameters. This energy region was chosen because
 it contains structures which bear relationships with these parameters,
 whereas the intermediate region 530--1560~eV was excluded because in
 this region measurement errors are greater than the errors attributable
 to the $t_j$ (see e.g.\ the abrupt changes in the difference curve at
690~eV and 900~eV, and note that the absolute deviation of the data from
 the model is much smaller at the lower energies where the transmittance
 is small). The result was $S_{19}=0.162$ and $S_{14}=0.160$,
$\Delta t_{\mathrm{Al}}= 0.27$~nm (for the original aluminium thickness
 before oxidation), $\Delta t_{\mathrm{pi}}=0.96$~nm, and
 $\Delta t_{\mathrm{ox}}=-0.56$~nm.
Next the correction of the grid thicknesses was sought that minimised
 the $L^1$ norm of the difference in the range 1560--1800~eV, yielding
$-0.26$~$\mu$m.
The residuals with the new window parameters are plotted in
 Fig.~\ref{fig:fit}c and their difference in Fig.~\ref{fig:fit}d.

The $\mu/\rho$ of aluminium was solved from Eq.~\ref{model} for both
 units, and the result is shown in Fig.~\ref{fig:al} for the two XAFS
 regions above the K and L absorption edges. The fine structures are
 typical of a solid, consisting of a multiple scattering peak in the
 narrow XANES (X-ray absorption near-edge structure) region just above
the absorption edge and oscillations from single scattering in the wider
 EXAFS (extended XAFS) region. Small differences between the two units
 are seen in the K-edge XAFS, and EXAFS clearly continues beyond
1800~eV\@. The results for carbon, nitrogen and oxygen are presented in
 Fig.~\ref{fig:cno}.  The difference in the oxide
 thickness between the two units may be
responsible for the apparent difference in the cross section of oxygen
(the oxide contributes 12\% to XAFS in unit~19 and 14\% in unit~14).

The remaining differences between the transmission data and the results
calculated from Eq.~\ref{model} using the extracted $\mu/\rho$ data are
 plotted in Fig.~\ref{fig:resid}. Data from the two units were averaged
 and interpolated in the appropriate energy ranges (carbon in
 283--400~eV, nitrogen in 400--530~eV, oxygen in 530--605~eV and
 aluminium in 73--282~eV and above 1554~eV) to yield the XAFS data for
 the modelling of the flight model window.
The K-edge EXAFS of aluminium was extrapolated into 1800--2000~eV with a
 third-order polynomial. The result is shown in Fig.~\ref{fig:extra}.
The resulting transmittance of the FM window is plotted in
 Fig.~\ref{fig:fm}.

\subsection{Analysis of line scans} \label{x-scan}

The line scans were analysed by solving a group of four nonlinear
 equations, constructed using Eq.~\ref{model} and the data at the four
 energies, at each beam position. The four variables were $S$ and the
 layer thicknesses, excluding the oxide which is the least absorbing.
 The results
 for the total aluminium thickness and for the polyimide membrane
 thickness are plotted in Fig.~\ref{fig:radial}. The overall radial
 variation is clearly explained by the total aluminium thickness which
decreases with the radial distance owing to the nature of the sputtering
process. Variations in the polyimide thickness do not exhibit any radial
 pattern.

As suggested by the curves in Fig.~\ref{fig:radial}, $t_{\mathrm{Al}}$
 can be modelled to good accuracy with the functional form
\begin{equation} \label{radial}
t_{\mathrm{Al}}(r) = \left[ 1 - \left( \frac{r}{r_0} \right)^2 \right]
		t_{\mathrm{Al}}(0) .
\end{equation}
This equation was fitted to the data with $r_0$, $t_{\mathrm{Al}}(0)$
 and the coordinates of the centre as free parameters. The best-fit
 $r_0$ was 77.74~mm for unit~19 and 77.56~mm for unit~14. The values for
 $t_{\mathrm{Al}}(0)$ were 60.43~nm (unit 19) and 59.61~nm (unit 14), in
 good accord with the $\Delta t_{\mathrm{Al}}$ of the energy scan
 analysis.

\subsection{Discussion} \label{chat}

The significance of the radial variation is modified by the point spread
function PSF of the SODART telescope. For an on-axis point source and
ideal alignment of SIXA with SODART, PSF decreases rapidly with small
values of $r$ and more slowly when $r$ increases, and the estimated
 fraction of photons encircled in the central element is about 0.6
 depending on the energy \cite{Finn}. The count rate at energy $E$ in a
 certain detector element is
\begin{equation} \label{phs}
C(E) = \int \Phi(h\nu) A_{\mathrm{eff}}(h\nu)
	    \left[ \int {\mathrm{PSF}}(h\nu,\vec{r}) \epsilon(h\nu,r)
	    \d a \right] P(h\nu,E) h \, \d \nu ,
\end{equation}
where $\Phi$ is the incident photon flux,
$A_{\mathrm{eff}}$ is the effective area of SODART,
 the surface integration is over the active area of the element
and $\epsilon$ is the detection efficiency
and $P$ the response function of the element.
Assuming perfect homogeneity of the crystal,
$\epsilon(h\nu,r)$ equals $T(h\nu,r)$ times a function of $h\nu$.

$T(r)$ with the radial dependence given by Eq.~\ref{radial} varies very
 slowly near the centre of the window. The surface integral over the
 central element lies between the two limits obtained by letting PSF
 approach a delta function and letting PSF be constant:
\begin{equation} \label{surf}
T(h\nu,0) \int \mathrm{PSF} \, \d a \ge
\int \mathrm{PSF}(h\nu,\vec{r}) T(h\nu,r) \d a \ge
\langle T \rangle \int \mathrm{PSF} \, \d a ,
\end{equation}
where the average of $T$ over the active area
 of the central element (radius $R$) is given by
\begin{equation}
\langle T \rangle = \frac{\int T \d a}{\int \d a} =
T(h\nu,0) \, \frac{\e^{FR^2} - 1}{FR^2} ,
\end{equation}
where
 $ F(h\nu) = {r_0}^{-2} t_{\mathrm{Al}}(0) \mu_{\mathrm{Al}}(h\nu) $.
The relative difference of the two limits is below $5\times 10^{-4}$
 in the energy range of SIXA; thus the first limit is a very good
 approximation for an on-axis point source.

For the other elements, the integration yields
\begin{equation} \label{outer}
\langle T \rangle = \frac{T(h\nu,0)}{FR^2}
  \left[ -1 + \e^{F D^2} \sum_{k=0}^\infty \frac{(FD)^{2k}}{(k!)^2} C_k
  \right] ,
\end{equation}
where $D$ is the distance between the centre of the element and the
 centre of the window, and
\begin{equation}
C_k = \left\{ \begin{array}{ll}
		\e^{FR^2} & \mbox{if } k=0 , \\
		R^{2k} \e^{FR^2} - \frac k F C_{k-1} & \mbox{if } k>0 .
	      \end{array}
      \right.
\end{equation}
The results calculated for the elements in the SIXA array are shown in
Fig.~\ref{fig:array}. The differences between the elements are at most
 1\% in the energy range of SIXA\@. When weighted with PSF as in the
 surface integral in Eq.~\ref{phs}, the averages become even closer to
$T(h\nu,0)$, because $T$ varies more slowly than PSF even for the outer
 elements. In any case, the smallness of the differences between the
 elements shows that Eq.~\ref{radial} with $r_0=77.7$~mm obtained by
 fitting to the experimental data from units 19 and 14 is a more than
 sufficiently accurate model of the positional variation of $T$ across
 the SIXA window.

The effect of XAFS on actual observations was studied by comparing data
 from a simulated observation to a model which neglects XAFS in the
 window materials. The simulated target was the Crab Nebula which was
 assumed to be a point source with a simple power-law spectrum
 ($\Phi \propto E^{-2.2}$) modified by interstellar absorption
 ($N_{\mathrm{H}}=3\times 10^{21}$~cm$^{-2}$).
This kind of spectrum with no emission or absorption line features is
 useful for revealing spurious lines arising from instrumental effects
 not properly accounted for in the response matrix. Both the simulated
data and the model were calculated for the central element of SIXA from
 Eq.~\ref{phs}, where the surface integral in Eq.~\ref{surf} was
 approximated by $0.6 \times T(h\nu,0)$ and $A_{\mathrm{eff}}$ was
 obtained from measurements with SODART telescope models \cite{priv}.
$T$ shown in Fig.~\ref{fig:fm} was used for the
 simulated data, while $T$ for the model was calculated from
 Eq.~\ref{model} with the data from ref.~\cite{Henke}.
The model by Scholze and Ulm \cite{Scholze2} was used for $P$ and the
 rest of $\epsilon$
 with experimental XAFS data for the K~edge of silicon \cite{Owens}.
Poisson noise, which results from the stochastic nature of the X-ray
 emission and detection processes, was added to the data assuming an
observation time of $5 \times 10^3$~s. The result of the simulation is
presented in Fig.~\ref{fig:crab}a. Spurious line features can hardly be
distinguished from Poisson noise, but if the noise was omitted from the
 calculation, it is observed that there is an emission line feature at
1.8~keV (equivalent width 1.5~eV, maximum residual 0.5\%), an absorption
 line at 0.6~keV (equivalent width 0.4~eV, maximum residual 0.3\%) and a
 continuous feature below the energy range of SIXA\@.

The smallness of the overall effect of XAFS and the similarity of XAFS
for the two window units suggest that the
derived XAFS is accurate enough for the flight model.
The main advantage of characterising the actual FM window would have
been the reduction of the uncertainty propagating from the uncertainties
 in the layer thicknesses and polyimide densities. The uncertainty of
the transmittance including the
propagated uncertainties of the layer thicknesses
is shown in Fig.~\ref{fig:fm}.
Uncertainties given in Table~\ref{tab:pi} were
 used for the polyimide membranes and grids. The mean deviation of the
 nominal value from the values of $t_{\mathrm{Al}}(0)$ derived
from the line scans (section \ref{x-scan}) was taken to be the
 standard deviation of
 $t_{\mathrm{Al}}$, while $\left| \Delta t_{\mathrm{ox}} \right|$
 (section \ref{e-scan}) was used for $t_{\mathrm{ox}}$.
The uncertainties of the derived
XAFS, based on the difference
in the data from the two window units, were negligible.
Uncertainties of $S$, the densities or the cross sections from
ref.~\cite{Henke} may have small contributions to the real uncertainty.

The significance
of the uncertainty is illustrated in Fig.~\ref{fig:crab}b with a
simulated case that the actual transmittance of the FM window is greater
 than modelled by the estimated uncertainty. The
 residual would then be 2.2\% at 0.5~keV and decrease towards higher
energies, falling below 1\% at 0.8~keV\@.

\section{Light transmission}

\subsection{Sensitivity of SIXA to light}

The photocurrent generated in a detector element is
\begin{equation}
I_{\mathrm{ph}} = qA \int \Phi(h\nu) T(h\nu) \eta(h\nu) h \d \nu ,
\end{equation}
where $\Phi$ is the incident flux,
 $A$ is the active area of the element and
$\eta(h\nu)$ is the quantum efficiency.
Fluctuations in the total current $I_{\mathrm{ph}} + I_{\mathrm{l}}$
(where $I_{\mathrm{l}}$ is the leakage current)
are a source of parallel
current noise. Expressed in terms of the number of carrier pairs,
the magnitude of the noise is
\begin{equation}
{\sigma_I}^2 = \sqrt{\frac{\overline{{N_{\mathrm{S}}}^2}}{q}
		     \left( I_{\mathrm{ph}} + I_{\mathrm{l}} \right)} ,
\end{equation}
where $\overline{{N_{\mathrm{S}}}^2}$ is a noise index which depends on
 the signal shaping
\cite{Goulding} (about 5~$\mu$s for SIXA).
The FWHM energy resolution is
\begin{equation}
\Delta E = \sqrt{(\Delta E_0)^2 +
		 (2.35 \, W)^2 \frac{\overline{{N_{\mathrm{S}}}^2}}{q}
		 I_{\mathrm{ph}}} ,
\end{equation}
where $\Delta E_0$ is the FWHM with no photocurrent
and $W$ is the mean pair-creation energy.

For example, the energy flux from a star of magnitude~5 is
$2.0 \times 10^5$ eV/(cm$^2$s). With an average photon energy of 2~eV,
the flux on the focal plane of SODART is then
 $1.1 \times 10^8$~cm$^{-2}$s$^{-1}$.
If all photons were absorbed in one element with 100\% efficiency,
the photocurrent without a window would be 12~pA
and the resolution would be degraded by 60~eV (from 200~eV).
If $I_{\mathrm{ph}}$ was reduced by two magnitudes with a 1\%
 transmitting window,
the resolution would increase by 0.7~eV only.

Silicon detectors can have very high quantum efficiencies for visible
light up to a cutoff wavelength which corresponds to the energy gap and
 is about 1.06~$\mu$m at the operating temperature of SIXA\@. For SIXA
crystals, $\eta(h\nu)$ depends mainly on the reflection and absorption
 properties of the cathode coating (gold--palladium alloy) on the
 detector surface. In addition, the incident flux is amplified by
 multiple reflections between the cathode and the window. We estimated
 $\eta(h\nu)$ with the help of a program which was written to compute
transmission properties of thin windows \cite{Harvela} as a part of the
 project {\em Development of thin optical filters and windows at XUV
wavelengths\/} which belongs to the General Support Technology Program
 of the European Space Agency.
The detector structure was approximated with a multilayer formed by
30~nm of gold, 3~mm of silicon and 250~nm of aluminium. The estimate of
 $\eta$ obtained by subtracting the absorptivity computed for a single
 gold layer from the absorptivity of the multilayer is presented in
Fig.~\ref{fig:qe}, together with $\eta$ multiplied by the amplification
 due to multiple reflections. The reflectances of the multilayer and of
 the window are also shown.

\subsection{Experimental}

The light transmission of SIXA window unit 16 was measured in the
 context of the mentioned ESA project. The measured data are plotted in
 Fig.~\ref{fig:light}a together with transmittance curves computed with
 the associated program. Interference peaks in the transmittance
computed using data of Table~\ref{tab:pi} for unit 16 (dotted curve) are
 shifted by 100--200~nm towards shorter wavelengths compared to the
 measured data. Interference wavelengths depend on the thickness of the
 polyimide membrane, but the shift does not necessarily imply a large
 error in the thickness measurement. Instead, the optical constants of
the thin film polyimide are probably different from those incorporated
 in the program because the material is  denser than the bulk material
and the polyimide is also of another type. With a 20\% greater thickness
 for the polyimide, the computed curves come closer to the data (dashed
 curves). The computed transmittance of the FM window is shown in
 Fig.~\ref{fig:light}b.

The transmission measurement and calculations suggest that the window is
more than sufficiently opaque.
As the worst case, Sirius could degrade the energy resolution of an
outer detector element by 0.01~eV, provided that its emission was
 concentrated in the region of maximum transmittance at 400--410~nm
 (where it actually has a strong H$\delta$ absorption line) and that all
 photons were absorbed in one element with 100\% efficiency. In the
 near-IR region the maximum transmittance between the interference
wavelengths of the two units is slightly higher. However, there are no
 astronomical sources outside the solar system with so high apparent
 magnitude in the near~IR that they could degrade the energy resolution
of SIXA\@.

The effect of photocurrent generated
 by light leakage through the window was also tested by comparing the
 resolution in the dark to the resolution when the detector was
illuminated with visible light. The detector used was a Si(Li) crystal
 produced for SIXA, connected to a preamplifier of a type similar to the
flight electronics. The crystal was operated in a cryostat with a mylar
window, and a piece cut from the SIXA window unit 16 was placed in front
 of the cryostat. The crystal was cooled with liquid nitrogen and heated
 to 122~K in order to generate some dark current. Filtered light from a
tungsten lamp was applied to generate photocurrent. The emission in the
wavelength range 250--800~nm was measured with a photodetector and it
 was of the order of $10^{-9}$~Wcm$^{-2}$nm$^{-1}$ for the longer
 wavelengths at a distance of 1.2~m.

When X-rays from a \nuc{55}{Fe} source were detected, the FWHM of the
 Mn~K$_\alpha$ peak was 187~eV in the dark. The interval between resets
 of the preamplifier was 12~s when the detector was not exposed to
 either light or X-rays. With the lamp switched on, it varied from 1.1~s
 to 30~ms with different filters used in front of the lamp. The reset
interval is directly proportional to $I_{\mathrm{ph}} + I_{\mathrm{l}}$,
 and fitting Eqs.~8 and 9 to the data yielded $I_{\mathrm{l}}=0.1$~pA\@.
The FWHM varied from 194 to 380~eV with the same filters. When another
 piece of the window unit 16 was added, no photocurrent was observed.
The incident spectra in the range 800--1060~nm were known poorly, but
 the measured photocurrents were of expected magnitude. The overall
 effective quantum efficiency for red and near-IR light was
roughly 30\%.

\section{Conclusions}

The X-ray entrance window for SIXA was characterised using synchrotron
radiation and two window units of a type similar to the flight model.
The contribution of the X-ray window to the system response of the SIXA
 spectrometer was adequately well determined. It can be accounted
for in the analysis of astronomical data by using the
energy and spatial dependence of the window transmittance
presented in this paper.

X-ray absorption fine structures of the constituting elements
were extracted from the data obtained with two non-flight window units.
The transmittance of the flight model window was calculated using the
derived cross sections in the regions where XAFS appears.

Spatial variation of the transmittance was observed to have a radial
 pattern which is attributed to a radial variation of the thickness of
 aluminium. The thickness could be closely modelled as a parabolic
 function of radial distance from the window centre. The variation is
less than 1\% in the energy range of SIXA\@.

Transmittance of visible and infrared light was studied and it was
 found that transmitted light from the observed targets
 will have no observable effect on the X-ray energy resolution.

\begin{ack}

We thank T. Lederer of the PTB for performing the SX700 measurements.
V.-P. Viitanen of Metorex International Oy is acknowledged for providing
the windows and answers to several questions about them.
 We thank also
 C. Budtz-J\o rgensen of the Danish Space Research Institute
(DSRI) for the beamtime and help at the SX700,
 H. Harvela of Metorex for the light transmission data
 and M.-A. Jantunen of Metorex
 and E. Tetri of the Helsinki University of Technology
 for collaboration with the photocurrent test.

\end{ack}

\newpage

\begin{figure}
\caption{Measured transmission at the centres of window units 19 (solid
 curve) and 14 (dotted curve). The insets show the absorption edges
 enlarged.}
\label{fig:e-scan}
\end{figure}

\begin{figure}
\caption{Measured transmission at different positions on the surfaces of
 window units~19 and 14 at four different X-ray energies. The $y$
 coordinate is 0 (solid curves), 10 (plus signs) or -10~mm (squares).}
\label{fig:x-scan}
\end{figure}

\begin{figure}
\caption{Adjustment of window unit parameters by the comparison of
 transmission data with the model: residuals (data/model - 1)
with (a) parameters from Table~1 and (c) adjusted parameters
for the window units 19 (solid curves) and 14 (dashed curves),
(b) the difference between the two residual curves in (a), and
(d) the difference between the two residual curves in (c).}
\label{fig:fit}
\end{figure}

\begin{figure}
\caption{Cross section of aluminium  above the K and L~absorption edges,
 extracted from the transmission data of window units
19 (solid curves) and 14 (dashed curves).
The dotted curves show the data from ref.~\protect\cite{Henke}.}
\label{fig:al}
\end{figure}

\begin{figure}
\caption{Cross sections of carbon, nitrogen and oxygen extracted from
the transmission data of window units 19 (solid curves) and 14 (dashed
 curves).
The dotted curves show the data from ref.~\protect\cite{Henke}.}
\label{fig:cno}
\end{figure}

\begin{figure}
\caption{Residuals between the measured and modelled transmittance for
 window units 19 (solid curve) and 14 (dashed curve).}
\label{fig:resid}
\end{figure}

\begin{figure}
\caption{Modelled linear attenuation coefficient
of aluminium derived from the
 transmission data between 1554--1800~eV and extrapolated into
 1800--2000~eV\@.
The dotted curve shows the data from ref.~\protect\cite{Henke}.}
\label{fig:extra}
\end{figure}

\begin{figure}
\caption{Calculated transmittance of the flight model window at its
centre. The upper curves are for units 6 and 17 and the lowest curve is
 the total transmittance. The insets show the uncertainty (standard
 deviation) of the transmittance calculated in section
 \protect\ref{chat}.}
\label{fig:fm}
\end{figure}

\begin{figure}
\caption{Variation of the total aluminium thickness
 $t_{\protect\mathrm{Al}}$ and the thickness $t_{\protect\mathrm{pi}}$
 of the polyimide membrane across the window surfaces. The residuals of
 $t_{\protect\mathrm{Al}}$ after fitting Eq.~\protect\ref{radial} are
 plotted in the central panels. The $y$ coordinate is 0 (solid curves),
 10 (plus signs) or -10~mm (squares).}
\label{fig:radial}
\end{figure}

\begin{figure}
\caption{Transmittance of the flight model window for the elements in
the SIXA array at different distances from the window centre. The curves
depict the difference of the transmittance averaged over the active area
of an element from the transmittance at the centre, and they apply (from
 top to bottom) to the outer 6 elements of the outer ring, inner 6
 elements of the outer ring, and 6 inner ring elements, respectively.}
\label{fig:array}
\end{figure}

\begin{figure}
\caption{Illustrating the accuracy of the window characterisation with
simulated observations of the Crab Nebula by SIXA:
(a) simulated data compared to an absorbed power-law model spectrum
 folded through an instrumental response matrix where
X-ray absorption fine structures in the window transmittance
are neglected, and
(b) assuming that the modelled transmittance of the window has an error
equal to the estimated standard deviation (Fig.~\protect\ref{fig:fm}).
In the upper panels the data are indicated by plus signs and the model
by solid curves. In the lower panels the plus signs indicate the
 residuals between the data and the model, while the solid curves
 represent the residuals with the data without Poisson noise.}
\label{fig:crab}
\end{figure}

\begin{figure}
\caption{Estimates of the quantum efficiency $\eta$ of SIXA detector
elements, the reflectances of the detector ($R_{\protect\mathrm{det}}$)
 and the window ($R_{\protect\mathrm{win}}$, computed for the centre of
 unit 6), and an "effective" quantum efficiency $\eta^*$ which
 incorporates the amplification of the incident flux by multiple
 reflections between the detector and the window.}
\label{fig:qe}
\end{figure}

\begin{figure}
\caption{Light transmittance of (a) window unit 16, measured (solid
 curves) and calculated (dashed and dotted curves), and
(b) flight model window (solid curves) and its constituting
units 6 (dashed curves) and 17 (dotted curves), calculated.
 The lower curves apply to the window centre and the upper curves
to the radial distance of 2.86~cm where the active detector area ends.
In order to compensate for the apparent errors in the optical constants
 of polyimide, 20\% greater thicknesses were used for the polyimide
 membranes in the computation of the dashed curves in (a) and all the
 curves in (b).}
\label{fig:light}
\end{figure}

\begin{table}[p]
\caption{Polyimide membrane and grid parameters of X-ray window units
 fabricated for SIXA\@. Units 6 and 17 have been chosen for the flight
 model, units 1 and 9 for the flight spare model and the others were
 used for calibration. The given $2\sigma$ uncertainties of the
 polyimide membrane thickness are uncertainties of the profilometer
 measurement multiplied by the correction which was applied to account
 for a systematic measurement error.}
\label{tab:pi}
\begin{tabular}{|c|c|c|c|c|}
\multicolumn{5}{c}{ } \\ \hline
Unit & Membrane & \multicolumn{3}{c|}{Grid} \\ \hline
 & thickness & thickness & width & shadowing \\ \hline
6 & 283 $\pm$ 17 nm & 17--18 $\mu$m & 42 $\mu$m & 16.1\% \\
17 & 308 $\pm$ 23 nm & 20--22 $\mu$m & 46 $\mu$m & 17.6\% \\ \hline
1 & 283 $\pm$ 17 nm & 19--20 $\mu$m & 44 $\mu$m & 16.8\% \\
9 & 268 $\pm$ 13 nm & 19--20 $\mu$m & 44 $\mu$m & 16.8\% \\ \hline
19 & 308 $\pm$ 23 nm & 19--21 $\mu$m & 46 $\mu$m & 17.6\% \\
14 & 308 $\pm$ 23 nm & 24--26 $\mu$m & 44 $\mu$m & 16.8\% \\
16 & 308 $\pm$ 23 nm & 20--21 $\mu$m & 44 $\mu$m & 16.8\% \\
 \hline
\end{tabular}
\end{table}

\end{document}